\documentclass[12pt]{article}
\usepackage[applemac]{inputenc}
\usepackage{amssymb,amsmath}
\usepackage{graphicx}
\usepackage[english]{babel}
\usepackage[square,comma,numbers,sort&compress]{natbib}

\begin{document}
\begin{titlepage}
\title{Comment on: ``Sadi Carnot on Carnot's theorem''.}

\author{ Jacques \textsc{Arnaud}
\thanks{Mas Liron, F30440 Saint Martial, France}, Laurent \textsc{Chusseau}
\thanks{ Centre d'\'Electronique et de Micro-opto\'electronique de Montpellier, Unit\'e Mixte de Recherche n°5507 au
CNRS, Universit\'e Montpellier II, F34095 Montpellier, France}, Fabrice \textsc{Philippe}
\thanks{D\'epartement de Math\'ematiques et Informatique Appliqu\'ees, Universit\'e Paul Val\'ery, F34199 Montpellier,
France.  Also with LIRMM, 161 rue Ada, F34392 Montpellier, France},}
\maketitle

\begin{abstract}      
Carnot established in 1824 that the efficiency $\eta_{C}$ of reversible engines operating between a hot bath at absolute
temperature $T_{hot}$ and a cold bath at temperature $T_{cold}$ is equal to $1-T_{cold}/T_{hot}$. Carnot particularly
considered air as a working fluid and small bath-temperature differences. Plugging into Carnot's expression modern
experimental values, exact agreement with modern Thermodynamics is found. However, in a recently published paper [``Sadi
Carnot on Carnot's theorem'', \textit{Am. J. Phys.} \textbf{70}(1), 42--47, 2002], G{\"u}{\'e}mez and others consider a
``modified cycle'' involving two isobars that they mistakenly attribute to Carnot. They calculate an efficiency
considerably lower than $\eta_{C}$ and suggest that Carnot made compensating errors. Our contention is that the Carnot
theory is, to the contrary, perfectly accurate.
\end{abstract}

%PACS 05.70.-a

\end{titlepage}

\section{Comment}

Carnot established in 1824 \citep{carnot} that the efficiency $\eta_{C}$ of reversible engines operating between a hot
bath at absolute temperature $T_{hot}$ and a cold bath at temperature $T_{cold}$ is equal to $1-T_{cold}/T_{hot}$. He
particularly considered air as a working fluid and small bath-temperature differences. If one plugs into Carnot's
expression modern experimental values one finds exact agreement with the above formula. A recently published paper by
G{\"u}{\'e}mez and others \citep{guemez} is useful in attracting attention to the early work on thermodynamics by
Carnot. However, the ``modified Carnot cycle'' that they consider, involving two isobars, is not the one treated by
Carnot. They calculate a considerably lower efficiency and suggest that Carnot made conceptual errors and employed
incorrect data, the agreement being achieved by coincidence only. To the contrary, our contention is that the Carnot
theory is perfectly accurate, and that his numerical estimates are fairly good. A related discussion was given in 1975
by \citet{hoyer}.

In his book ``R{\'e}flexions sur la puissance motrice du feu'', published in 1824, \citet{carnot} presents calculations on
the ``motive power’’ of heat engines, defined as ``the useful effect that an engine is capable of producing. The effect
can always be expressed in terms of a weight being raised to a certain height. It is measured by the product of the
weight and the height to which the weight is considered to have to be raised''. Specifically, Carnot employed as the
mechanical energy unit the energy required to lift a cubic meter of water by one meter in the earth gravitational field,
that is, 9.81 kJ. As far as heat consumption is concerned, Carnot employed as a unit the heat required to raise one
kilogram of water from 0 to 1°C (say, at constant pressure). Thus, the Carnot unit for heat is equal to 4.18 kJ. 

\citet[p. 80]{carnot} considers an ideal heat engine whose working agent is a cylinder containing 1 kg of air initially
at atmospheric pressure $p$=10.4 meters of water. The cold bath temperature is $T_{cold}$=0°C while the hot bath
temperature $T_{hot}$=1°C \footnote{Initially, Carnot considered a hot-bath temperature of 0.001 °C, but later on
switched to 1°C.}. Let us postpone physical explanations and consider the values of the work $W$ performed per cycle and
the hot-bath heat consumption $Q$ as given in Carnot's book: 
\begin{eqnarray}
\label{W}
W=\Delta V \Delta p \qquad \Delta V=(\frac{1}{116}+\frac{1}{267})\;0.77\qquad \Delta
p=\frac{10.4}{267} \nonumber\\ Q=0.267\Longrightarrow \frac{W}{Q}=0.00138\; \text{Carnot units}.
\end{eqnarray}

If we introduce into the above Carnot formula recent experimental values and convert heat into energy, we have instead
\begin{eqnarray}
\label{Wbis}
W=(\frac{1}{109.3}+\frac{1}{273.15})\;0.773\;\frac{10.34}{273.15} \nonumber\\
Q=0.240\Longrightarrow \eta=\frac{W}{Q}\frac{9.81}{4.18}=0.00367.
\end{eqnarray}
in nearly exact agreement with the Carnot efficiency $\eta_{C}= 1-T_{cold}/T_{hot}\approx 1/273.15=0.00366$.

\begin{figure}[hbpt]
\begin{center}
\includegraphics[scale=0.75]{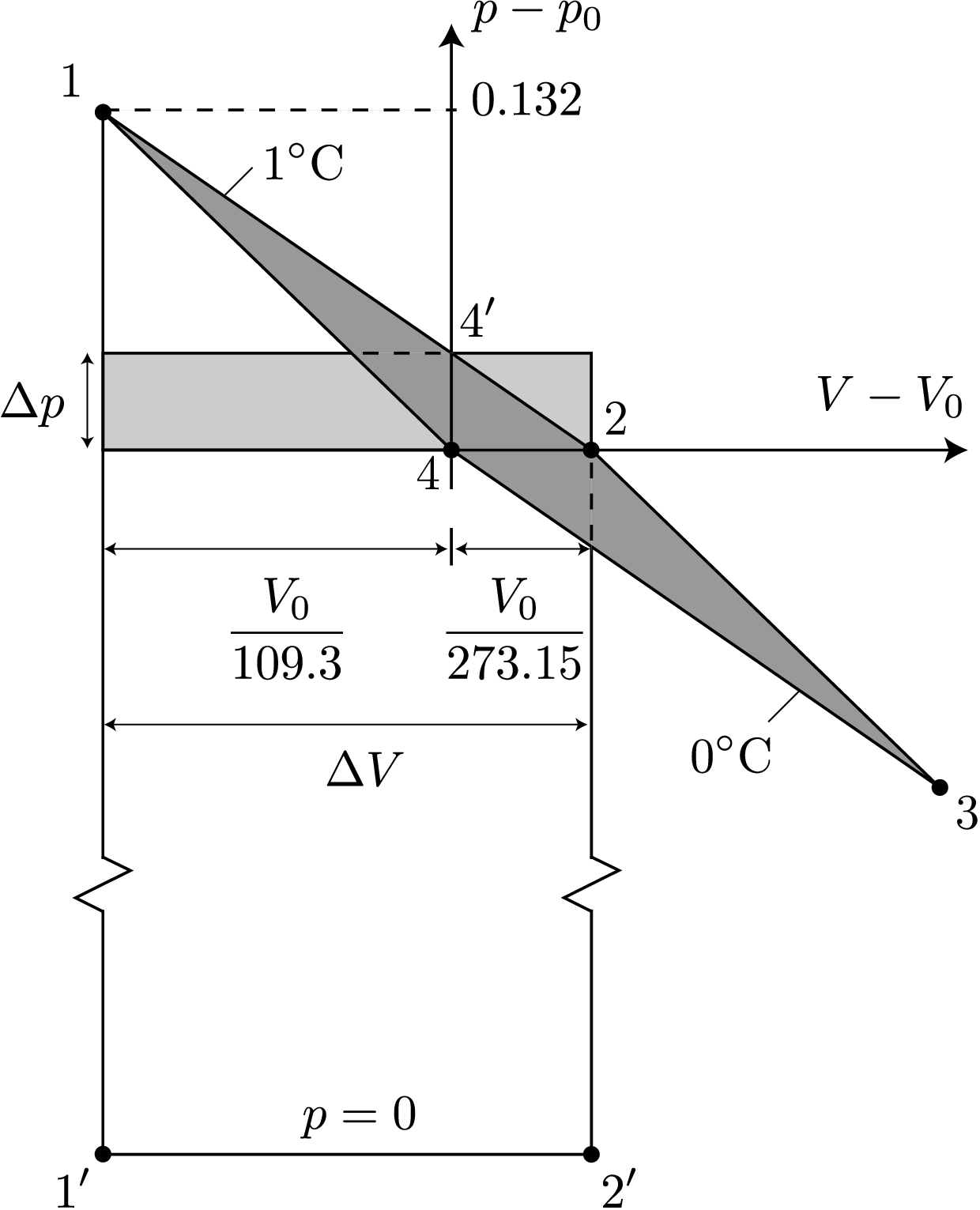}
\caption{This figure represents at scale the reversible cycle ($1$-$2$-$3$-$4$-$1$) for air, considered by Carnot.
Recent experimental values are shown. At the origin of the diagram ($T=0$°C) the pressure is $p_{0}$=10.34 meters of
water and the volume is $V_{0}=0.773$ cubic meters. The work performed is equal to the area enclosed in the
paralelogram, also equal to the area enclosed in the rectangle, namely: $W=\Delta V ~\Delta p$. The heat consumption $Q$
is approximately equal to the heat required to heat air from 0°C to 1°C at atmospheric pressure. The cycle efficiency
$\eta \approx 1/273.15$. \label{carnotfig}}
\end{center}
\end{figure}

The reasoning that led Carnot to the expression in \eqref{W} is sound. Figure \ref{carnotfig} shows the reversible cycle
$1$-$2$-$3$-$4$-$1$ considered by Carnot in the pressure-volume diagram. The vertical axis corresponds to the change of
pressure in the air-filled cylinder with respect to atmospheric pressure $p_{0}$, while the horizontal axis corresponds
to the change in cylinder volume with respect to the volume at atmospheric pressure and $T=0$°C, namely $V_{0}=0.773$
cubic meters. Because the relative changes of temperature and volume are small the cycle is a parallelogram. For obvious
geometrical reasons the work performed per cycle, that is, the area enclosed in the parallelogram, is: $W=\Delta p \;
\Delta V$, where $\Delta p$ and $\Delta V$ are shown in the figure. This is also the area enclosed in the rectangular
path shown in the figure. Let us emphasize, however, that this path is \emph{not} the cycle considered by Carnot. 

$\Delta p$ is the change of pressure required to increase the temperature of 1 kg of air from 0 to 1°C, the volume being
kept constant (path $4$-$4'$). This quantity had been measured at Carnot's time by Gay-Lussac. The recent value is $\Delta
p=p_{0}/273.15$, where $p_{0}=10.34$ meters of water. Carnot introduced the ratio $\gamma$ of air specific heats at
constant pressure and constant volume, which is also the ratio of the isothermal and isentropic compressibilities (see,
for example, \citep[p. 272]{zemansky}). According to the figure, the ratio of the slopes of the isotherms and adiabats
is $\gamma=(1/109.3+1/273.15)/(1/109.3)=1.400=7/5$. This is indeed the ratio of constant-pressure to constant-volume
heat capacities of di-atomic molecules such as those comprising air (oxygen and nitrogen) in the temperature range
considered. A slightly larger value of $\gamma$ was used by Carnot on the basis of the sound-velocity measurements made
at the time. Thus the volume change $\Delta V=(1/109.3+1/273.15) \; V_{0}$, where $V_{0}=0.773$ cubic meters.

To obtain the amount of heat $Q$ supplied by the hot bath, Carnot noted that for small temperature differences the work
performed is negligible compared to the heat consumption. It follows that one may assume, for that part of the
calculation, that the heat received by the system almost vanishes in a closed cycle. Consider now the cycle
$1$-$2$-$4$-$1$ (triangular path). By definition, no heat is being transfered to the fluid along the adiabat $4$-$1$.
Therefore, the amount of heat $Q$ received by the system along the path $1$-$2$ is equal to the heat received along the path
$4$-$2$. The latter is the heat required to raise the temperature of 1 kg of air from 0 to 1°C at (constant) atmospheric
pressure. Experimental values for this quantity were known at Carnot's time.

Carnot also correctly noted that, for small temperature differences, essentially the same values of $W$ and $Q$ are
obtained if the cylinder volume is kept constant when being transfered from one bath to the other, that is, adiabats may
be replaced by isochores, in that limit. However, when the cylinder is kept in contact with a bath and its volume
changes as shown in the figure, the pressure necessarily varies. This is why the description of the Carnot cycle given
in \citep{carnot} ``The cycle used by Carnot was composed of two isobarics and two isochorics'' is erroneous. It is then
not surprising that these authors calculate a much lower efficiency, namely, $\eta=0.00089$, using modern data, and
misinterpret the nature of Carnot's contribution, as far as air or other nearly ideal gases are concerned.

Carnot gave in a manuscript the mechanical equivalent of heat according to: 1 kilo-calorie of heat=370 kg.m (instead of
the modern value of 426 kg.m corresponding to 1 calorie of heat=4.18 J). Because the energy of ideal gases depends on
temperature only, $Q$ is equal to the work $\Delta V~ p_{0}$ done along the isothermal path $1$-$2$. If we replace $Q$
in \eqref{W} by $\Delta V~  p_{0}$, $\Delta V$ drops out and we obtain, using the ideal gas law $pV/T=$constant, $\eta =
\Delta p / p_{0}=1-T_{cold}/T_{hot}$.

Readers interested in the history of Thermodynamics would do well in reading the La Mer papers (reference 6 of the
commented paper) that appeared long ago in this journal.
 
\bibliographystyle{unsrtnat}
\bibliography{%
CarnotComment%
}

\end{document}